\documentclass{article}
\usepackage{spconf,amsmath,amssymb,graphicx,hyperref}

\title{HINT: Hierarchical Inter-frame Correlation for One-Shot Point Cloud Sequence Compression}
\name{Yuchen Gao, Qi Zhang\thanks{This research was supported by the TOAST project, funded by the European Union’s Horizon Europe research and innovation program under the Marie Skłodowska-Curie Actions Doctoral Network (Grant Agreement No. 101073465) and NordForsk Nordic University Cooperation on Edge Intelligence (Grant No. 168043).\\
Authors' e-mails: \{yuchen, qz\}@ece.au.dk.\\
}
}
\address{DIGIT and Department of Electrical and Computer Engineering, \\
Aarhus University, Aarhus, Denmark\\
}
\begin{document}
\ninept
\maketitle
\begin{abstract}
Deep learning has demonstrated strong capability in compressing point clouds. Within this area, entropy modeling for lossless compression is widely investigated. However, most methods rely solely on parent/sibling contexts and level-wise autoregression, which suffers from decoding latency on the order of $10^1-10^2$ seconds. We propose \textbf{HINT}, a method that integrates temporal and spatial correlation for sequential point cloud compression. Specifically, it first uses a two-stage temporal feature extraction: (i) a parent-level existence map and (ii) a child-level neighborhood lookup in the previous frame. These cues are fused with the spatial features via element-wise addition and encoded with a group-wise strategy. Experimental results show that HINT achieves encoding and decoding time at 105\,ms and 140\,ms, respectively, equivalent to 49.6\,$\times$ and 21.6\,$\times$ acceleration in comparison with G-PCC, while achieving up to 43.6\% bitrate reduction and consistently outperforming the spatial-only baseline (RENO).   
\end{abstract}
\section{Introduction}
\label{sec:intro}
Volumetric 3D content such as LiDAR scans and human body captures is rapidly proliferating in autonomous driving, AR/VR, and telepresence. However, raw point clouds are extremely bandwidth-hungry, which makes efficient geometry compression a prerequisite for scalable capture, transmission, and storage. The MPEG standards provide two major toolsets: geometry-based point cloud compression (G-PCC) and video-based point cloud compression (V-PCC)~\cite{gpcc,vpcc}. While robust and widely deployed, these methods still face limitations in compression efficiency and do not fully exploit the intricate structure of point cloud sets.
 \begin{figure}[htbp]
    \centering
    \includegraphics[width=0.95\linewidth, trim=.2cm 1.7cm .2cm 0.2cm, clip]{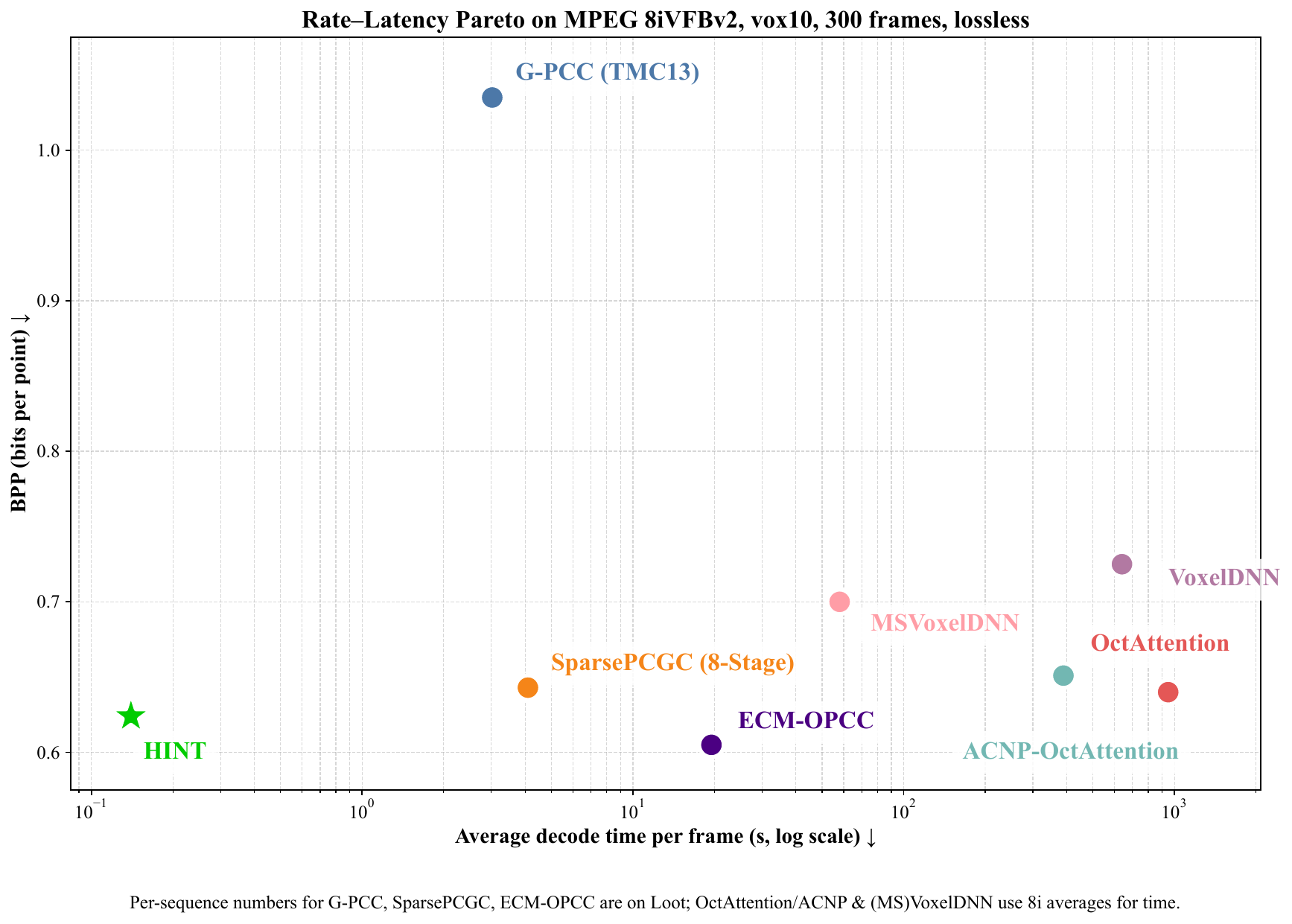}  
    \caption{Rate-latency Pareto on MPEG 8iVFBv2.}
    \label{fig:rl-compare}
\end{figure}  
 
Recently, learning-based methods have shown strong capability in entropy modeling that estimates the probability of octree/voxel occupancy for arithmetic coding. Entropy models are attractive for their causal decoding and standard-friendly bitstream integration, but they are often designed per-frame and thus do not fully exploit the temporal redundancy. We aim to bridge the gap between fast, strictly causal entropy models and temporally aware coding for point cloud sequences.
A comparison of rate-latency in Fig.~\ref{fig:rl-compare} provides an illustration that HINT not only achieves high compression ratio but also accelerates encoding/decoding.  

\begin{figure*}[htbp]
    \centering

    \includegraphics[width=\textwidth,trim=0cm 0.4cm 0.7cm 0cm, clip]{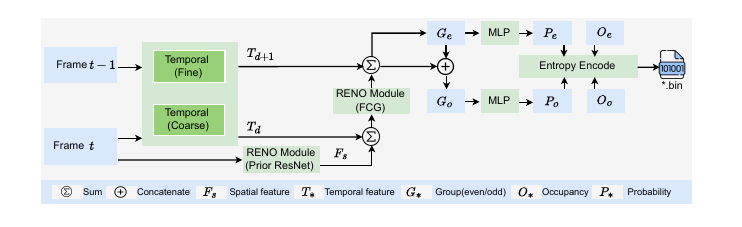}  
   \caption{Overview of HINT. Frame $t$ and $t-1$ are processed by the Temporal Module to produce a parent-level feature $T_{d}$ and a child-level feature $T_{d+1}$. The RENO path provides spatial feature $F_s$ from frame $t$ (Prior ResNet at parents). The parent-level features are fused by an element-wise addition and broadcast to child level. Then, it is fused with $T_{d+1}$ as the final feature. Each parent's eight children are split by parity into two groups, and we first predict $G_e$'s probability with the final feature. Ground truth $G_e$ is embedded and aggregated with final feature as additional information to predict $G_o$'s probability. The predicted probabilities are passed to an entropy encoder to produce the bitstream.}

    \label{fig:pipeline}
\end{figure*}  
Our proposed \textbf{HINT} is an entropy modeling method that augments spatial contexts with temporal correlation derived from the previously decoded frame to improve compression. We design a coarse-to-fine extraction module to track changes from frame $t\!-\!1$ to $t$. Then we fuse temporal and spatial correlation to predict occupancy symbol distributions. The encoding/decoding voxels are separated into groups to enhance the context for estimation. Experimental results (Sec.\ref{sec:exp}) demonstrate that HINT significantly accelerates encoding/decoding compared with other methods and consistently outperforms spatial-only baseline (RENO) in compression ratio.

\section{RELATED WORK AND BACKGROUND}
\label{sec:related}
\noindent\textbf{Standards.}~G-PCC employs octree subdivision with transforms and predictive coding~\cite{gpcc}, while V-PCC projects to 2D patches and reuses video codecs~\cite{vpcc}. MPEG’s G-PCC and V-PCC remain the most widely deployed baselines but depend on hand-crafted octree/transform tools or 2D projections, which leaves spatial structure and temporal redundancy to be explored in many scenarios. Early “differential octree’’~\cite{kammerl2012real} has pointed out the potential in temporal redundancy for dynamic point cloud streaming. While these standards are robust and widely adopted, their tools (tree transforms, prediction along Morton order, 2D patch projection) are not optimized for modern GPU-parallel inference and often leave fine-grained 3D structure and short-range temporal redundancy under-exploited. This gap motivates learned entropy models that keep native 3D structure and run efficiently on parallel hardware.

\noindent\textbf{From standards to learned codecs.}~With advances in deep learning and GPU hardware, learning-based codecs have been proven to outperform the standard approaches both in latency and compression ratio. Octree/voxel entropy models~\cite{huang2020octsqueeze,que2021voxelcontext,nguyen2021multiscale} model occupancy with multi-level priors, which reduces bits, but decoding is still sequential. Most cues come from parents/ancestors, offering limited sibling interaction. Transformer-based octree codecs~\cite{jin_ecm-opcc_2023, fu_octattention_2022,sun_enhancing_2024} address this by grouping children and using self-attention to learn richer sibling context. However, they still rely on autoregression across groups and levels, and the attention blocks add non-trivial compute and memory, so practical decoders often run in seconds per frame. Most learned codecs still focus on spatial contexts within a single frame. Temporal modeling either relies on explicit motion estimation or autoregression, which hurts latency. Our design keeps strict causality and introduces lightweight temporal and sibling cues that are naturally parallelizable.

 \begin{figure*}[htbp]
  \centering
  \includegraphics[width=\linewidth, trim=.2cm 0.3cm .2cm 0.2cm, clip]{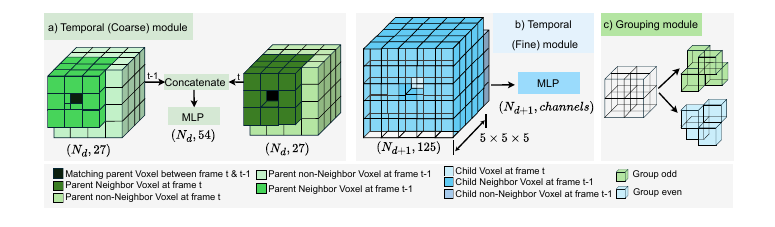} 
    \caption{Temporal context and grouping module. a) Temporal (coarse) module. For each parent-level voxel at time $t$, we collect a size $V_d$ (e.g., $V_d$=27) neighborhood at frames $(t,t\!-\!1)$, concatenate the retrieved occupancy from the two frames. b) Temporal (fine) module. For each child-level voxel at frame $t$, we query a window at the corresponding location in frame $t\!-\!1$, feed it to an MLP to obtain a per-child temporal feature. c) Grouping. The children of each parent are split into odd and even groups.}
  \label{fig:coarse}
\end{figure*}

Among entropy estimation methods, RENO sidesteps octree construction by operating in a multiscale sparse-tensor space~\cite{wang_sparse_2022} and compressing sparse occupancy codes at each level~\cite{you_reno_2025}. Specifically, dyadic down/up-scaling yields parent-child relations without building explicit trees. To achieve fast down/up-scaling: (i) Fast Occupancy Generator (\textsc{FOG}), a fixed-weight sparse convolution module derives child occupancy codes from coordinates, and (ii) Fast Coordinates Generator: (\textsc{FCG}) reconstructs child coordinates from parent coordinates and their occupancy code (8-bit code in $[0,255]$) (two non-learned operations). Let $\mathcal{C}_d$ be the coordinate set at any level $d$, and $\mathcal{O}_d$ be the occupancy codes.~RENO expresses the geometry tensor as: ${\mathcal{C}}_d \;\equiv\; (((\mathcal{C}_0,\mathcal{O}_0),\mathcal{O}_1),\ldots,\mathcal{O}_{d-1}).$ Hence, lossless compression aims to compress the input point cloud with maximum level $D$ by: 
\vspace{-1mm}
\begin{equation}
\mathcal{C}_D=(((\mathcal{C}_0,\mathcal{O}_0),\mathcal{O}_1),\ldots,\mathcal{O}_{D-1}).
\end{equation}

Here $\mathcal C_0$ denotes root coordinates and $\mathcal O_0$ corresponds to the root occupancy code. According to arithmetic coding~\cite{rissanen1979arithmetic}, we need the occupancy symbol distribution $P^{(d)}$ to encode the occupancy code at any level $d$. However, the ground truth $P^{(d)}$ is unknown. RENO estimates a $\hat{P}_\theta^{(d)}$ under the condition of parent occupancy status:
\begin{equation}
\begin{aligned}
\hat P_\theta(\mathcal O_1,\dots,\mathcal{O}_{D-1})
= \prod_{d=1}^{D-1}
\hat P_\theta^{(d)}\!\big(\mathcal{O}_d \mid \mathcal{O}_{d-1},\, \mathcal{C}_{d-1}\big).
\end{aligned}
\label{eq:prob_spatial}
\end{equation}
A two-stage ResNet (Prior ResNet) encodes the parent occupancy code $\mathcal O_{d-1}$ and propagates the feature by another module (Target ResNet) to its children and outputs child-level features. Each 8-bit occupancy code $O_d$ is split into $(s_1, s_0)$, where $s_0$ are the least-significant four bits and $s_1$ the most-significant four bits. Then, $s_0$'s probability is estimated by an MLP with the child-level features. The ground truth $s_0 $ is embedded with child-level features. They are fed into another MLP to estimate the probability of $s_1$. According to Shannon's theory~\cite{shannon1948}, the expected code length per symbol is $-\log_2 P(\cdot)$. Thus, for the joint distribution over $\mathcal O=(\mathcal O_1,\ldots,\mathcal O_{D-1})$, we learn model parameters $\theta$ by minimizing the cross-entropy (to reduce the expected code length):
\begin{equation}
\begin{aligned}   
\theta &\xleftarrow{}
\arg\min_\theta \mathbb{E}_{\mathcal O \sim P}\!\left[-\log_{2}\hat P_\theta(\mathcal O)\right]\\
&= \sum_{d=1}^{D-1}\mathbb{E}_{\mathcal O_d\sim P}\!\left[
-\log_{2}\hat P_\theta^{(d)}\!\big(\mathcal{O}_d \mid \mathcal O_{d-1},\mathcal C_{d-1}\big)
\right].
\end{aligned}
\end{equation}
RENO conditions each child purely on spatial contexts within one frame, without explicit temporal modeling or sibling context. This leaves temporal correlations and sibling dependencies under-exploited, which our method addresses with lightweight temporal cues and parity-grouped sibling conditioning. Furthermore, RENO is assessed only on sparse LiDAR sequences. We extend it to high-density point cloud sequences.
\vspace{-3mm}

 \section{HINT}
 
HINT consists of three primary steps: (i) optional quantization and hierarchy construction \footnote{A standard approach to process the raw point cloud, but for MPEG 8i Voxelized Full Bodies v2 (8iVFBv2) we skip this step as it has been quantized already.} , (ii) temporal and spatial feature extraction, and (iii) entropy coding using estimated probability.
 
\subsection{Hierarchical sparse representation}
For a quantized frame, we need to preprocess the input point cloud and build the hierarchy. We construct a multi-level sparse pyramid by leveraging RENO's FOG module to acquire the coordinates and occupancy code for each resolution level. At level $d$ with $N_d$ voxels, we obtain a sparse tensor $(\mathcal{C}_d, \mathcal{O}_d)$, where $\mathcal{C}_d\in\mathbb{Z}^{N_d\times 4}$ stores the integer coordinates and $\mathcal  O_d\in\{0,\ldots,255\}^{N_d}$ contains the occupancy code.
 
\subsection{Parent-level temporal correlation}
\label{sec:coarse}
We define current coding level as child-level $d+1$, its corresponding parent-level is at $d$. To integrate temporal correlation, we operate at these two levels. For any child voxel at $d+1$, we find its parent voxel at $d$ and build an ``existence map" between frame $t$ and $t-1$ (Fig.~\ref{fig:coarse}(a)). For each current parent-level coordinate $C_{d}^{(t)}\in \mathcal{C}_d^{(t)}$, we define a cubic offset set $\mathcal{N}_{{d}}$ of size $V_{d}$ ($V_{d}{=}7$ for face-adjacent neighbors, $V_{d}{=}27$ for $3{\times}3{\times}3$ or $V_{d}{=}125$ for $5{\times}5{\times}5$). $V_{d}$ is a tunable hyperparameter, we use $V_{d}$=27 unless otherwise stated. We query existence at $C_{d}^{(t)}+\delta$ for all $\delta\in\mathcal{N}_{d}$ in both frames:
\begin{equation}
\mathbf{z}^{(t)}(C_{d}^{(t)}+\delta),\ \mathbf{z}^{(t-1)}(C_{d}^{(t)}+\delta)\in\{0,1\}^{V_d},
\end{equation}
where elements in $\mathbf{z}$ indicate the existence at $C_{d}^{(t)}+\delta$ ($1$ if present, else $0$). This operation is computed by one neighborhood expansion and two binary searches over sorted Morton keys of $C_{d}^{(t)}$ and $C_{d}^{(t-1)}$.
The ``existence map" $ M $ is produced by concatenation as follows: \begin{equation}
M=\operatorname*{concat}_{dim=1}[\mathbf{z}^{(t)}(C_{d}^{(t)}+\delta)\ ,\mathbf{z}^{(t-1)}(C_{d}^{(t)}+\delta)].\end{equation} 
Then, $ M$ is transformed by an MLP 
 to a feature with a dimension of 32 channels. Specifically, we map the $2V_{d}$-dim vector to a 32-dim feature $T_{d}$ and this feature is additively fused with the RENO produced spatial feature $F_s$. The fused feature is broadcast by: 
 \begin{equation}
      F_d = \operatorname{FCG}(F_s+T_{d}),
 \end{equation}
where $F_d$ is the per-child feature. If the coarse level feature remains stable across frames, child voxels at $d{+}1$ tend to preserve the previous frame’s local occupancy pattern. Otherwise, the model downweights the temporal cue.
 
 
\subsection{Child-level temporal correlation}
\label{sec:fine}
The fine level temporal module is designed to complement the coarse level output. At level $d+1$, we estimate frame $t$ child occupancy codes based on the occupancy pattern from frame $t-1$ (Fig.~\ref{fig:coarse}(b)). 

We define a query offset set $\mathcal{N}_{d+1}$ of size $V_{d+1}$ (default $V_{d+1}=125$, i.e., a window of $5\times5\times5$). For each child coordinate $C^{(t)}_{d+1}\!\in\!\mathcal{C}_{d+1}^{(t)}$, we traverse all
$\delta\in\mathcal{N}_{d+1}$ and look up $C^{(t)}_{d+1}+\delta$ in the previous-frame set
$\mathcal{C}_{d+1}^{(t-1)}$. If found, we read its 8-bit occupancy code
$O^{(t-1)}_{d+1}(C_{d+1}^{(t)}+\delta)\in\{0,\ldots,255\}$, otherwise $0$ for absent. Let $E\in\mathbb{R}^{256\times32}$ be a 256-entry embedding table. For each offset, we take the embedding of occupancy code $O^{(t-1)}_{d+1}$ at $C^{(t)}_{d+1}+\delta$ by $e_\delta=E\!\big(O^{(t-1)}_{d+1}(C^{(t)}_{d+1}+\delta)\big)$. We average the neighbor descriptors over $\mathcal{N}_{d+1}$, and apply a linear projection $W_{t}$ to obtain a 32-dim feature:
\begin{equation}
    T_{d+1}= W_{t}\cdot \frac{1}{|\mathcal{N}_{d+1}|}\sum_{\delta\in\mathcal{N}_{d+1}} e_\delta,
\end{equation}
where $W_{t}\in\mathbb{R}^{32\times32}$ is a learnable matrix. Finally, we fuse the temporal cue $T_{d+1}$ with the broadcast coarse feature $F_d$ to acquire the final feature $F_{d+1}$ by element-wise addition:
\begin{equation}
F_{d+1}=F_d+T_{d+1}.
\end{equation}






\vspace{-3mm}

\begin{table*}[htbp]
\centering

\small
\caption{Mean BPP (bits per point) over 300 frames (vox10). (\textendash{} denotes not reported in the original paper.)}

\label{tab:bpp_compare}
\resizebox{\textwidth}{!}{%
\begin{tabular}{l|cccccccc}
\hline
\textbf{Dataset} & \textbf{G-PCC} & \textbf{OctAttention} & \textbf{ACNP-OctAttention} & \textbf{ECM-OPCC} &\textbf{SparsePCGC}&\textbf{VoxelDNN}&\textbf{MSVoxelDNN}&\textbf{HINT}\\
\hline
\emph{Loot} & 0.970 & 0.620  &  0.596    & 0.550 &  0.596    &  0.580  &  0.730&   0.562   \\

\emph{Soldier}&1.030 &\textendash{} &\textendash{}&\textendash{} &0.628 &\textendash{}&\textendash{} &0.582\\
\emph{Longdress} &1.015 &\textendash{}&\textendash{}&\textendash{}& 0.625& 0.670&\textendash{}& 0.604\\
\emph{Redandblack} &1.100 &0.660 &0.706 &0.660 &0.690&0.870 &0.670& 0.686\\
\hline

\end{tabular}}

\end{table*}
\vspace{-6mm}

\begin{table*}[t]
\centering
\small
\caption{Computational complexity comparison on \emph{Loot} \& \emph{Redandblack} (mean per frame). (\textendash{} denotes not reported in the original paper.)}
\label{tab:complexity_cols}
\begin{tabular}{l|cccc}
\hline
 & \textbf{G-PCC} & \textbf{SparsePCGC}&\textbf{ECM-OPCC} & \textbf{HINT} \\
\hline
Model size (MB)    & \textendash{}                 & 4.9 & 8.0                 & \textbf{3.4} \\
Encoding time (ms) & 5215               & 3799 &  1920                 & \textbf{105}  \\
Decoding time (ms) & 3023               & 4095 &     19500             & \textbf{140}  \\
\hline
\end{tabular}
  
\end{table*} 

\subsection{Entropy coding based on sibling correlation}
\label{sec:group}
 To leverage sibling correlation while preserving causality, a grouping strategy to enhance the context is designed (Fig.~\ref{fig:coarse}(c)). We split each parent’s eight children into two groups: $G_{e}=\{0,3,5,6\}$, $G_{o}=\{1,2,4,7\}$ \footnote{Index $i$ for child voxel is written as $(b_x,b_y,b_z)\in\{0,1\}^3$. If $b_x{+}b_y{+}b_z$ is
even (e.g., $3=(1,1,0)$, sum$=2$) then $i\in G_e$, otherwise $i\in G_o$
(e.g., $1=(1,0,0)$, sum$=1$). Flipping one axis bit toggles the parity, so the
three face neighbors of any child lie in the opposite group.}.
 
 We first follow RENO's two-stage bitwise coding for $G_e$. For each child voxel $i\in G_e$ at level $d+1$, let the 8-bit occupancy code be split into $O_{d+1}(i)=(s_1(i),s_0(i))$, where $s_0(i)$ is
the least-significant $4$ bits and $s_1(i)$ the most-significant $4$ bits. We first predict $P(s_0(i)\!\mid\!F_{d+1})$ and encode $s_0(i)$. We then embed $s_0(i)$ (via a 16-entry table) and add it to $F_{d+1}$ to predict $P(s_1(i)\!\mid\!F_{d+1},s_0(i))$. Arithmetic coding is applied after each prediction.

Before encoding $G_o$, we summarize the already decoded even siblings into a lightweight context vector.
Let $O_e(i)\in\{0,\ldots,255\}$ be the decoded 8-bit occupancy code for child $i\in G_e$. We embed each code with a 256-entry table $E\in\mathbb{R}^{256\times 32}$ and concatenate with its 3-dim relative position $\pi(i)\in\mathbb{R}^3$ (the child’s $(b_x,b_y,b_z)$ inside the parent), then we apply a linear projection $W_{s}\in\mathbb{R}^{32\times(32+3)}$. The four descriptors are averaged with a mask (over available siblings) to form a sibling feature of the same width as $F_{d+1}$:
 \begin{equation}
     F(G_e)=\frac{1}{|G_e|}\sum_{i\in G_e} W_{s}\,[E(O_e(i)),\,\pi(i)],
 \end{equation}
 where $E(\cdot)$ is a 256-entry embedding table. For any child voxel $j\in G_o$, we fuse this context by element-wise addition for richer context $\tilde F_{d+1}$:
$  
 \tilde F_{d+1}=F_{d+1}+ F(G_e).
 $
Then we reuse the same two prediction heads:
 first predict $s_0(j)$ from $\tilde F_{d+1}$, then embed $s_0(j)$,
add it to $\tilde F_{d+1}$, and predict $s_1(j)$. Arithmetic coding is
applied after each prediction. This keeps strict causality: odd children depend
only on already decoded even siblings within the same parent.
\vspace{-2mm}

\section{Experiments}
\vspace{-2mm}

\label{sec:exp}
All experiments are run on a desktop with an Intel Core i9-14900 CPU and an NVIDIA RTX-4090 GPU. We implement the codec in PyTorch~\cite{paszke2019pytorch} with TorchSparse~\cite{tang2022torchsparse} for sparse 3D operations. We use the torchac~\cite{mentzer2019practical} for entropy encoding. 

\noindent\textbf{Dataset} We evaluate the MPEG 8i Voxelized Full Bodies v2 (8iVFBv2) sequences at 10-bit quantization across 300 frames: \emph{Loot}, \emph{Longdress}, \emph{Redandblack}, and \emph{Soldier}~\cite{d2019jpeg}. Unless stated, results are lossless coding. For each sequence we report the mean bits-per-point (BPP) over 300 frames.

\noindent\textbf{Baseline} We compare against standard G-PCC~\cite{gpcc} and representative learning-based entropy models that target octree/voxel occupancy: OctAttention~\cite{fu_octattention_2022}, ECM-OPCC~\cite{jin_ecm-opcc_2023}, ACNP-OctAttention~\cite{sun_enhancing_2024}, SparsePCGC~\cite{wang_multiscale_2021}, VoxelDNN~\cite{nguyen2021multiscale}, and MSVoxelDNN~\cite{nguyen2021multiscale}. We also include RENO~\cite{you_reno_2025} as a spatial correlation-only baseline.

\noindent\textbf{Main results} Tab.~\ref{tab:bpp_compare} summarizes mean BPP at 10-bit precision on 8iVFBv2. 
HINT is competitive across all sequences: it achieves comparable compression ratio to ECM-OPCC and consistently outperforms G-PCC, SparsePCGC, OctAttention, ACNP-OctAttention, and VoxelDNN. 
Although with a slight rate gap to ECM-OPCC, Tab.~\ref{tab:complexity_cols} shows that HINT reduces encoding and decoding latency by one to two orders of magnitude compared with ECM-OPCC. 
This combination of compression and computational efficiency highlights HINT as a codec for practical deployment. Tab.~\ref{tab:complexity_cols} shows HINT runs at 105/140\,ms (encode/decode) ($36.2\times$/$29.3\times$ speedups over SparsePCGC and $49.6\times$/ $21.6\times$ over G-PCC), with a 3.4\,MB (vs.\ 4.9\,MB for SparsePCGC) model size.
\begin{figure}[!htb]
  \centering
  \includegraphics[width=\linewidth, trim=2.2mm 1mm 2.2mm 7.5mm, clip]{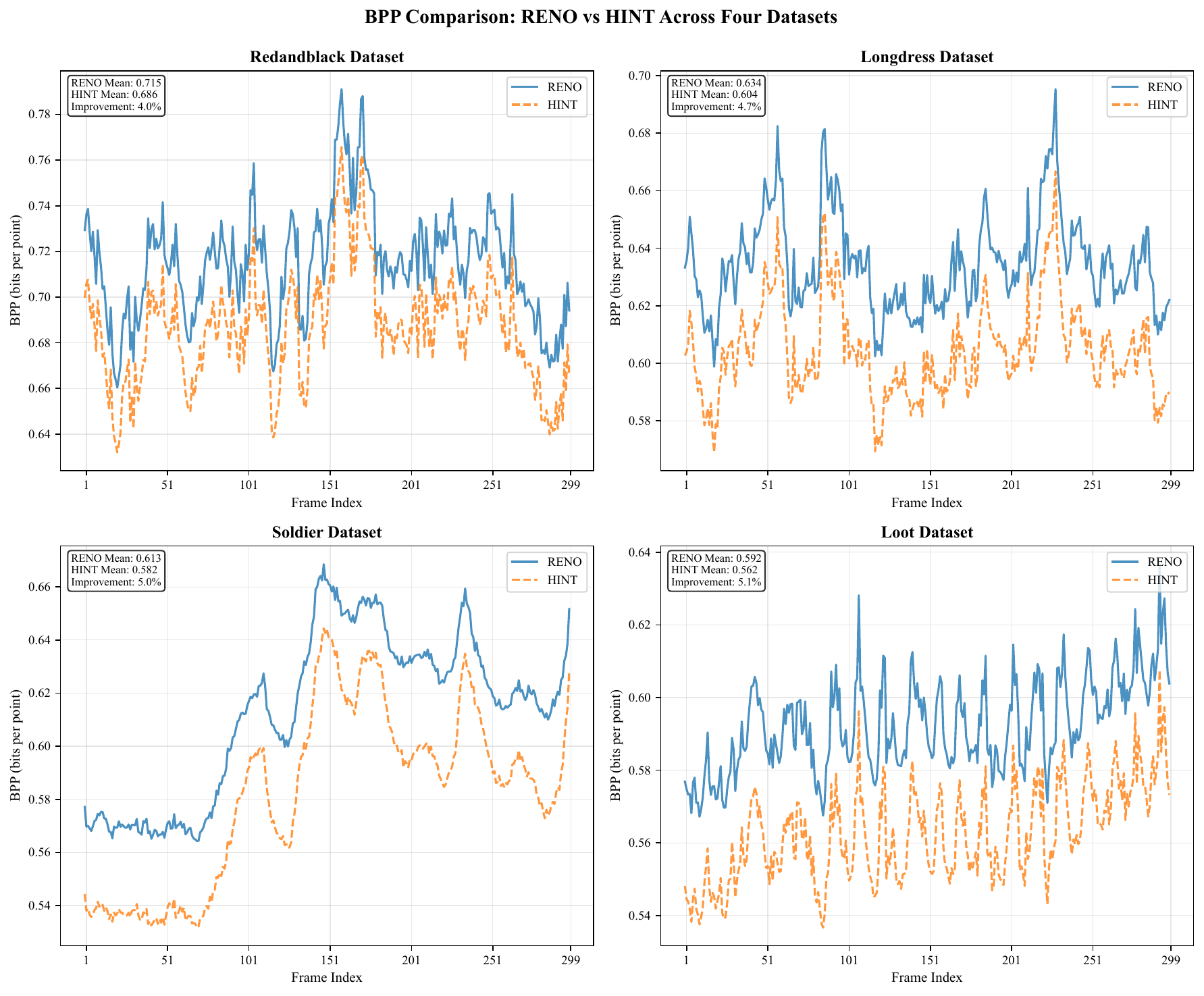}
\vspace{-5mm}
  \caption{BPP comparison on MPEG 8iVFBv2 (vox10).}
  \label{fig:compare_with_reno}
\end{figure} 
Fig.~\ref{fig:compare_with_reno} plots per-frame BPP against RENO on the four sequences (\emph{Redandblack}, \emph{Longdress}, \emph{Soldier}, and \emph{Loot}). 
When consecutive frames are relatively static, more inter-frame correlation exists and HINT achieves 4-5\% average savings (up to 6.7\% in \emph{Soldier}). When consecutive frames differ significantly, less correlation can be exploited and the benefit diminishes. However, HINT's performance never falls below spatial correlation-only baseline RENO, ensuring consistent improvements across time.

\noindent \textbf{Ablation (Loot, mean over 300 frames).}
Removing temporal module increases BPP from 0.562 to 0.568, while removing sibling conditioning increases it to 0.587 (RENO: 0.592), suggesting sibling conditioning contributes more to the bitrate reduction. Meanwhile, temporal cues provide alignment-free gains, consistent with our low-latency design.

\vspace{-1mm}
\section{Conclusion}
 
In this work, we introduce HINT, a dynamic point cloud sequence compressor that leverages both a spatial entropy model and temporal cues. A coarse parent-level existence map and a fine child-level neighborhood lookup are fused through an MLP and paired in groups for encoding/decoding. In 8iVFBv2 dataset, HINT not only reduces the mean bitrate (e.g., 4 to 5\% vs.\ RENO and up to 43.6\% vs.\ G-PCC), but also achieves 49.6$\times$ and 21.6$\times$ speedups in encoding and decoding over G-PCC. These results suggest that temporal and sibling redundancy can be exploited without explicit motion estimation or heavy autoregression. However, temporal gains may diminish under weak inter-frame coherence (abrupt motion/occlusion/noise). Future work will explore confidence-gated and adaptive temporal conditioning without motion estimation.

\clearpage
\bibliographystyle{IEEEbib}
\bibliography{strings,refs}

\end{document}